\documentstyle[12pt,epsf]{article}
\begin{document}
\title{Stability of 3$D$ Cubic Fixed Point in Two-Coupling-Constant
$\phi^4$-Theory}
\author{H.\ Kleinert, S.\ Thoms and V.\ Schulte-Frohlinde\\
Institut f\"ur Theoretische Physik \\
Freie Universit\"at Berlin \\
Arnimallee 14 \\
14195 Berlin}
\date{ }
\maketitle
\begin{abstract}
For an anisotropic euclidean $\phi^4$-theory with two interactions
$[u (\sum_{i=1}^M {\phi}_i^2)^2+v \sum_{i=1}^M \phi_i^4]$
the $\beta$-functions are calculated from five-loop perturbation expansions
in $d=4-\varepsilon$ dimensions,
using the knowledge of the large-order behavior and Borel transformations.
For $\varepsilon=1$, an infrared stable cubic
fixed point for $M \geq 3$ is found,
implying that the critical exponents in the magnetic phase transition
of real crystals are of the cubic universality class.
There were previous indications of the stability based either on
lower-loop expansions or on less reliable Pad\'{e} approximations,
but only the evidence presented in this work seems to be
sufficently convincing to draw this conclusion.
\end{abstract}
\section{Introduction}
The most elegant approach to phase transitions in many physical 
systems proceeds via
field-theoretic renormalization group techniques \cite{CriPhe}.
The best-studied system is the isotropic Heisenberg ferromagnet with $M$
classical spin components. Its critical behavior can be described
correctly by an O($M$)-symmetric vector field theory with a quartic
interaction $\sum_{i=1}^M \left({\phi}_i^2\right)^2$.

In a real crystal, such an interaction is never presentably by itself.
The crystalline structure gives rise to anisotropies,
most prominently of cubic symmetry, which can be represented by an
extra field interaction $\sum_{i=1}^M {\phi}_i^4$.
This term breaks the O($M$)-symmetry by favoring magnetizations
along the edges or the diagonals of a hypercube in $M$ dimensions.
The extended theory interpolates between an O($M$)-symmetric
and a cubic system.
It has been pointed out a long time ago~\cite{Aharon}
that, depending on $M$, the O($M$)-symmetric and the
cubic fixed point interchange their stability.
For $M < M_c$, the O($M$)-symmetric, {\em isotropic} fixed point is stable.
For $M > M_c$ the isotropy is destabilized and the trajectories
of renormalization flow cross over to the {\em cubic} fixed point.
Estimates using calculations up to three-loops \cite{Aharon}-\cite{BreGuZ1}
indicated that $M_c$ must lie somewhere between $3$ and $4$.
Resummation procedures based on Pad\'{e} approximations \cite{MaSo}
suggested $M_c$ to lie below $3$, thus permitting real crystals to exhibit
critical exponents of the cubic universality class.
The uncertainty of these estimates have prompted Kleinert and
Schulte-Frohlinde \cite{KleSchu2} to carry the expansions up to
five loops. They increased the evidence for $M_c < 3$ considerably,
again via Pad\'{e} resummation.

For a simple $\phi^4$-theory, the Pad\'{e} approximation is known to be
inaccurate.
At present, the most accurate renormalization group 
functions for that theory
have been obtained by combining perturbation expansions
with large-order estimates, using a resummation procedure based on
Borel-transformations  \cite{ResPro}-\cite{GuZin5}.

Intending the application of these more powerful resummation methods,
the large-order behavior of renormalization group functions
has recently been derived for the $M$-vector model with cubic anisotropy,
by Kleinert and Thoms \cite{KleTho}.

It is the purpose of this letter to 
combine these large-order results with the
five-loop perturbation expansions \cite{KleSchu2} using
a simple Borel-type of resummation algorithm \cite{ResPro,KleJan},
whose power has been exhibited in recent model studies by Kleinert,
Thoms and Janke \cite{KThoJa}.

The results to be presented in this paper allow us to conclude 
with a reasonable certainty that
an infrared-stable cubic fixed point exists at the physically most relevant
value $M=3$. However, due to the vicinity of the isotropic fixed point, the
differences in the critical exponents are very hard to measure experimentally.
Going beyond the Pad\'{e} work in Refs. \cite{MaSo} and \cite{KleSchu2},
we also show explicitly the instability and stability of the isotropic and 
the cubic fixed point, respectively.
\section{Resummation}
\subsection{The Problem}
Object of investigation is a ${\phi}^4$-theory with cubic
anisotropy. The corresponding energy functional reads:
\begin{equation}
\label{Hamilt}
\hspace*{-0.2cm}
H(\vec{\phi})=\displaystyle{\int}\!d^dx
\left[\frac{1}{2} {\partial}_{\mu}{\phi}_{Bi} {\partial}_{\mu}{\phi}_{Bi}+
\frac{8 {\pi}^2}{3} \left(\frac{u_B}{4} S_{ijkl}+
\frac{v_B}{4} {\delta}_{ijkl}\right) {\phi}_{Bi}{\phi}_{Bj}
{\phi}_{Bk}{\phi}_{Bl}\right]  ,
\end{equation}
where ${\phi}_{Bi}(x)\ (i=1,2, \ldots ,M)$ is the bare $M$-component field
in $d=4-\varepsilon$ dimensions, and $u_B$, $v_B$ are the bare coupling
constants. In particular, we shall consider the physically most interesting
case of $M=3$, and continue $\varepsilon$ to $\varepsilon=1$.
The tensors associated with the two interaction terms in (\ref{Hamilt}) have
the following symmetrized form:
\begin{eqnarray}
S_{ijkl}\!\!\!&=&\!\!\! \frac{1}{3} (\delta_{ij} \delta_{kl}+
                        \delta_{ik} \delta_{jl}+
                        \delta_{il} \delta_{jk})\, ,\nonumber\\
\delta_{ijkl}\!\!\!&=&\!\!\! \left\{ \begin{array}{l}
                        1\, , \quad i=j=k=l\, , \\ 0\, , \quad {\rm otherwise}\, .
                                    \end{array} \right.
\nonumber
\end{eqnarray}
The symmetry of the action under reflection ${\phi}_i \rightarrow -{\phi}_i$
and under permutations of the $M$ field indices $i$ implies the following
form of the vertex functions, persisting to all orders in perturbation theory:
\begin{equation}
{\Gamma}^{(2)}_{ij} \sim {\Gamma}^{(2)} {\delta}_{ij}\, ; \quad
{\Gamma}^{(2,1)}_{ij} \sim {\Gamma}^{(2,1)} {\delta}_{ij}\, ,
\end{equation}
\begin{equation}
{\Gamma}^{(4)}_{ijkl} \sim {\Gamma}^{(4)}_u S_{ijkl}+
{\Gamma}^{(4)}_v \delta_{ijkl} \, .
\end{equation}
The symmetry permits us to renormalize the
$M$ components of the field $\vec{\phi}$ and the
composite field $\frac{1}{2} {\vec{\phi}}^2$ with only a
single renormalization constant $Z_{\phi}$ and $Z_{{\phi}^2}$, respectively.
The bare field $\vec{\phi}_B$, the composite field
$\frac{1}{2} {\vec{\phi}}^2_B$, and the two coupling
constants $u_B$ and $v_B$ are  related to the
corresponding physical objects by
\begin{equation}
\label{RGconst}
\begin{array}{rclrcl}
\vec{\phi}_B (x)\!\!\!&=&\!\!\!Z_{\phi}^{1/2}\vec{\phi}(x) \, ;&\quad
\left[{\vec{\phi}}^2_B\right] (x)\!\!\!&=&\!\!\!(Z_{{\phi}^2})^{-1}
\left[{\vec{\phi}}^2 \right](x) \, , \\
u_B\!\!\!&=&\!\!\!{\mu}^{\varepsilon} Z_u (Z_{\phi})^{-2} u \, ;&\quad
v_B\!\!\!&=&\!\!\!{\mu}^{\varepsilon} Z_v (Z_{\phi})^{-2} v \, ,
\end{array}
\end{equation}
where $\mu$ is a mass parameter.
We employ dimensional regularization with minimal subtraction.
The square brackets around ${\vec{\phi}}^2$ indicate a renormalization
of this composite operator. Recall that this renormalization is different
from that of the wave function which converts
$\vec{\phi}_B$ to $\vec{\phi}$. In fact it is closely related to the mass  
renormalization of a theory with mass $m \neq 0$. In the minimal subtraction
scheme, the corresponding renormalization constants are 
related by: $Z_{{\phi}^2}=Z_{m^2}(Z_{\phi})^{-1}$.
 
The RG-functions are defined in the usual way:
\begin{equation}
\label{RGfunct}
\begin{array}{cclccl}
{\beta}^u(u,v)\!\!\!&=&\!\!\!\mu {\partial}_{\mu}u
\left|_{u_B,v_B,\varepsilon}\right. \, ;&\quad
{\beta}^v(u,v)\!\!\!&=&\!\!\!\mu {\partial}_{\mu}v
\left|_{u_B,v_B,\varepsilon}\right. \, ,\\
{\gamma}_{\phi}(u,v)\!\!\!&=&\!\!\!\mu {\partial}_{\mu}\ln Z_{\phi}
\left|_{u_B,v_B,\varepsilon}\right. \, ;& \quad
{\gamma}_{{\phi}^2}(u,v)\!\!\!&=&\!\!\!-\mu {\partial}_{\mu}\ln Z_{{\phi}^2}
\left|_{u_B,v_B,\varepsilon}\right. \, .
\end{array}
\end{equation}
The natural parameter for the anisotropy of the system is the ratio
$\delta=v/(u+v)$, and the isotropic case corresponds to $\delta=0$.
We shall use the new couplings $g=u+v$ and $\delta$ for the
calculation of the fixed points from the resummed $\beta$-functions
\begin{eqnarray}
\label{Beta}
{\beta}^g(g,\delta)\!\!\!&=&\!\!\!{\beta}^v[u(g,\delta),v(g,\delta)]+
{\beta}^u[u(g,\delta),v(g,\delta)] \, ,
\nonumber\\
g {\beta}^{\delta}(g,\delta)\!\!\!&=&\!\!\!(1-\delta){\beta}^v[u(g,\delta),
v(g,\delta)]-\delta {\beta}^u[u(g,\delta),v(g,\delta)] \, .
\end{eqnarray}
The O($M$)-symmetric and cubic fixed points can be obtained
by calculating the simultaneous zeros $(g^{\ast},{\delta}^{\ast})$
of ${\beta}^u[u(g,\delta),v(g,\delta)]$
and ${\beta}^v[u(g,\delta),v(g,\delta)]$.
For the physically interesting number of field components, $M=3$,
the infrared-stable cubic fixed point is expected to appear very close
to the O($M$)-symmetric one. Since $\delta$ is very small in this region,
it will be sufficient to restrict the resummation efforts to 
the $g$-series accompanying each power ${\delta}^n$, so that
the $\beta$-functions at the cubic fixed point will be approximated by
\begin{eqnarray}
\label{zeros}
0\!\!\!&=&\!\!\!{\beta}^u[u(g^{\ast},{\delta}^{\ast}),v(g^{\ast},
{\delta}^{\ast})]\approx\sum_{n=0}^N B_n^{u(N)}(g^{\ast}){\delta}^{
\ast \,n}\, , \nonumber \\
0\!\!\!&=&\!\!\!{\beta}^v[u(g^{\ast},{\delta}^{\ast}),v(g^{\ast},
{\delta}^{\ast})]\approx{\delta}^{\ast}\sum_{n=1}^N B^{v(N)}_n(g^{
\ast}){\delta}^{\ast \,n-1} \, ,
\end{eqnarray}
where $B^{(N)}_n(g) \equiv {\rm res} \left[\sum_{k=n}^N \beta_{kn} g^k\right]$
indicates resummed $g$-series. 

From the five-loop perturbation expansion
in Ref. \cite{KleSchu2}, the perturbation coefficients ${\beta}^{u/v}_{kn}$
are known up to the order $N=6$. For $\varepsilon=1$ and the number of
field components $M=3$, the following expansions are known:
\begin{eqnarray}
\sum_{k=0}^6 {\beta}^u_{k0} g^k \!\!\! &=&\!\!\! -g+3.667 g^2-7.667 g^3+
47.651 g^4-437.646 g^5+4998.62 g^6 \, ,
\nonumber \\
\sum_{k=1}^6 {\beta}^u_{k1} g^k \!\!\! &=&\!\!\! g - 5.333 g^2+15.667 g^3-
121.767 g^4+1341.05 g^5-17821.1 g^6 \, ,
\nonumber \\
\sum_{k=2}^6 {\beta}^u_{k2} g^k\!\!\! &=&\!\!\! 1.667 g^2-10 g^3+115.885 g^4-
1664.86 g^5+27191 g^6 \, ,
\nonumber \\
\sum_{k=3}^6 {\beta}^u_{k3} g^k\!\!\! &=&\!\!\! 2 g^3-50.074 g^4+1064.62 g^5-
22916.2 g^6 \, ,
\nonumber \\
\sum_{k=4}^6 {\beta}^u_{k4} g^k\!\!\! &=&\!\!\! 8.305 g^4-350.528 g^5+
11183.1 g^6
\, , \nonumber \\
\sum_{k=5}^6 {\beta}^u_{k5} g^k\!\!\! &=&\!\!\! 47.368 g^5-2966.14 g^6 \, ,
\quad {\beta}^u_{66} g^6=330.76 g^6
\, , \nonumber \\
& & \nonumber \\
\mbox{and} \ \ \ \ \ \ \  & &  \nonumber \\
& & \nonumber \\
\sum_{k=1}^6 {\beta}^v_{k1} g^k\!\!\!&=&\!\!\! -g+4 g^2-10.778 g^3+75.875 g^4-
776.26 g^5+9707.36 g^6
\, , \nonumber \\
\sum_{k=2}^6 {\beta}^v_{k2} g^k\!\!\!&=&\!\!\! -g^2+6.222 g^3-67.319 g^4+
944.05 g^5-15030.9 g^6
\, , \nonumber \\
\sum_{k=3}^6 {\beta}^v_{k3} g^k\!\!\!&=&\!\!\! -1.111 g^3+30.211 g^4-
639.243 g^5+13549.6 g^6
\, , \nonumber \\
\sum_{k=4}^6 {\beta}^v_{k4} g^k\!\!\!&=&\!\!\! -6.218 g^4+233.262 g^5-
7122.94 g^6
\, , \nonumber \\
\sum_{k=5}^6 {\beta}^v_{k5} g^k\!\!\!&=&\!\!\! -33.414 g^5+1973.58 g^6
\, , \quad {\beta}^v_{66} g^6=-228.19 g^6 \, .
\end{eqnarray}
In addition to the five-loop expansions, the
large-order behavior of ${\beta}^u_{kn}$ and ${\beta}^v_{kn}$ has been
obtained explicitly in Ref. \cite{KleTho}, with the result
\begin{equation}
\label{betaLO}
\beta^{u/v}_{kn}\stackrel{k \rightarrow \infty}{\longrightarrow}\,
 {\gamma}^{u/v}(n) \left(-1\right)^k k!\, k^{(d+5)/2+n} \left[
 1+{\cal O}(1/k)\right]\, ,\quad k \gg n \, .
\end{equation}
The resummation algorithm to be employed in this paper will make use of all
these informations.
\subsection{The Algorithm}
As explained in detail in Ref. \cite{KThoJa},
it is possible to reexpand a divergent perturbation series
\begin{equation}
\label{Zseries}
Z(g,\delta)=\sum_{k=0}^{\infty}\sum_{n=0}^k Z_{kn} \, g^k {\delta}^n \, .
\end{equation}
in a special infinite set of Borel summable functions $I_{pn}(g)$ as
\begin{equation}
\label{Zresum}
Z(g,\delta)= \sum_{n=0}^{\infty}
\left[\sum_{p=n}^{\infty} a_{pn}\, I_{pn}(g)\right]
{\delta}^n\,,
\end{equation}
so that the approximation
\begin{equation}
\label{Zapprox}
Z(g,\delta) \approx Z^{(N)}(g,\delta)= \sum_{n=0}^N
\left[\sum_{p=n}^N a_{pn}\, I_{pn}(g)\right]
{\delta}^n=\sum_{n=0}^N Z^{(N)}_n(g) {\delta}^n
\end{equation}
has the same series as $Z(g,\delta)$ up to the powers $g^N {\delta}^N$,
while reproducing
the known large-order behavior of the perturbation expansion (\ref{Zseries}).
\begin{equation}
\label{ZLO}
Z_{kn}\stackrel{k \rightarrow \infty}{\longrightarrow}\,
 {\gamma}(n) \left(-\sigma \right)^k k!\, k^{{\beta}_n} \left[
1+{\cal O}(1/k)\right]\, , \quad k \gg n \, .
\end{equation}
As shown by Janke and Kleinert in Refs. \cite{ResPro,KleJan},
the natural choice for the functions $I_{pn}(g)$ are certain confluent
hypergeometric functions,
where for the case of two coupling constants a second index was introduced
in Ref. \cite{KThoJa}.
One possibility was to use
\begin{eqnarray}
\label{ZIpn}
\lefteqn{I_{pn}(g)=}  \\
& &\left(\frac{4}{\sigma g}\right)^{b_0(n)+1}\int_0^1 dw
   \frac{(1+w) w^{b_0(n)+p}}{\Gamma\left[b_0(n)
   +1\right](1-w)^{2b_0(n)+2\alpha+3}}
   \exp\left[-\frac{4w}{(1-w)^2 \sigma g}\right] \nonumber
\end{eqnarray}
with
\begin{equation}
\label{bo}
b_0(n)={\beta}_n+\frac{3}{2} \, .
\end{equation}
Then the coefficients $a_{pn}$ are given by
\begin{equation}
\label{apn}
a_{pn}=\sum_{k=n}^p \frac{Z_{kn}}{(b_0(n)+1)_k}
\left(\frac{4}{\sigma}\right)^k
\left(\begin{array}{c}
        p+k-1-2\alpha \\ p-k
       \end{array} \right)\, ,
\end{equation}
where $c_k=\Gamma(c+k)/{\Gamma(c)}$\ are Pochhammer's symbols.
The parameter $\alpha$ is free to choose, and may be used
to accommodate any strong-coupling power of $Z(g,\delta)$
\begin{equation}
\label{Zstrong}
Z(g,\delta)\stackrel{g \rightarrow \infty}{\longrightarrow}\,
\kappa(\delta) g^{\alpha} \, ,
\end{equation}
if this is known.
Since in quantum field theory, this is not the case, $\alpha$ will be chosen 
by the condition of best convergence of the resummed
$g$-series $Z^{(N)}_n(g)$ in Eq. (\ref{Zapprox}),
as explained in the next section.
\subsection{Optimal Choice of Strong-Coupling Power $\alpha$}
\begin{figure}[p]
\unitlength1cm
\vspace{-8ex}
\begin{minipage}[t]{13.2cm}
\epsfxsize=13cm
\hfill
\epsfbox{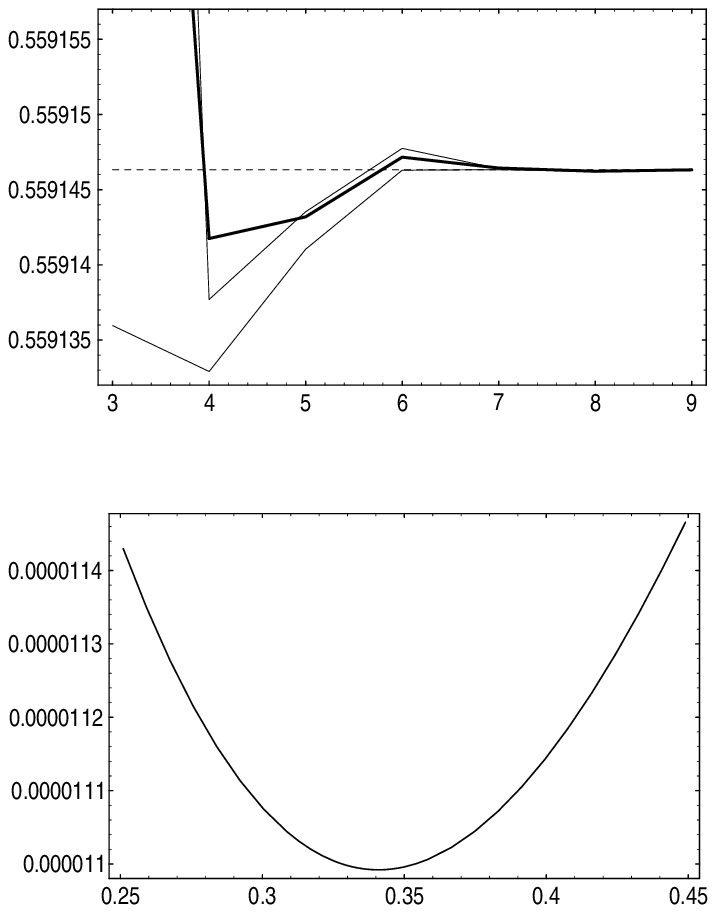} \par
\begin{picture}(13,2)
\put(1.7,13.5){$E^{(N)}$}
\put(2.5,12.6){{\footnotesize $E_{\rm exact}$}}
\put(1.8,5.7){$\Delta^{(7)}$}
\put(9.5,14.0){$g/4$ = $0.1$ }
\put(6,6.8){${\alpha}_{\rm exact}$ = $1/3$}
\put(6,6.2){${\alpha}^{\Delta}_{\rm PMS}$ = $0.3408$ }
\put(4,14.0){$a)$}
\put(4.3,7.5){$b)$}
\linethickness{0.05mm}\put(7.5,11.75){\line(-1,0){2.0}}
\linethickness{0.04mm}\put(7.5,11.1){\line(-5,1){1.8}}
\linethickness{0.05mm}\put(7.5,10.5){\line(-1,0){2.0}}
\put(7.8,11.75){${\alpha}_{\rm exact}$ = $1/3$}
\put(7.8,11.1){$\alpha$ = $0.55$}
\put(7.8,10.5){$\alpha$ = $0.15$}
\put(7.5,8.85){$N$}
\put(7.5,2.25){$\alpha$}
\end{picture}
\end{minipage}
\vspace{-10ex}  
\caption[]{\label{conv}
Convergence of the ground-state energy $E$ of the anharmonic
oscillator with $g/4=0.1$. $a)$ Resummed ground-state energies $E^{(N)}$ are
plotted versus the order of approximation {\it N} for various
strong-coupling parameters $\alpha$. $b)$ The $\alpha$-dependence of the 
function $\Delta^{(7)}$ of Eq. (\ref{Epms}).
The optimal $\alpha$-value is given by the 
minimum ${\alpha}^{\Delta}_{\rm PMS}=0.3408$, very close to the exactly
known value $1/3$. }
\end{figure}

An idea of the relevance of $\alpha$  is gained
from the study of corresponding models in quantum mechanics, where the
strong-coupling behavior can be deduced from scaling arguments.
Consider first an O($M$)-symmetric anharmonic oscillator with
$g (x_i^2)^2/4$ interaction, where $M$ 
is the number of components of the
vector $\vec{x}$. Here the functions $Z^{(N)}_0(g)$ in Eq. (\ref{Zapprox})
represent the resummed ground-state energies $E^{(N)}(g)$.
In Figure \ref{conv} we have illustrated the convergence of $E^{(N)}$
for the anharmonic oscillator with one $x$-component $(M=1)$
at a coupling constant $g/4=0.1$. We have plotted $E^{(N)}$ versus the 
order of approximation $N$ for various values of
the strong-coupling parameter $\alpha$.
At large $N$, the curves become increasingly independent of $\alpha$ and
approach a saturation value which coincides with the ground-state energy.
This exact result does not depend on $\alpha$. Therefore, we
choose the strong-coupling parameter under the condition that the curvature
and the slope of the corresponding curve depend minimally on
the variation of $\alpha$, when approaching the saturation region.
This choice complies with the
{\em principle of minimal sensitivity} (PMS) which has been
used with great success to optimize
variational perturbation expansions~\cite{KleiPI}.
Using the discretized form of the first and second derivative,
the optimal $\alpha$-value, to be denoted by ${\alpha}^{\Delta}_{\rm PMS}$, 
is found by calculating the extrema or turning 
points of the judicial function
\begin{eqnarray}
\label{Epms}
\Delta^{(N_{\rm s})}(g,\alpha)&=& \\ 
& &\!\!\!\!\!\!\!\!\!\!\!\!\!\!\!\!\!\!\!\!\!\!\!\!\!\!\!\!\!\!\!\!\!
\left[(E^{(N_{\rm s}-1)}-E^{(N_{\rm s}-2)})^2+
(E^{(N_{\rm s})}-2 E^{(N_{\rm s}-1)}+E^{(N_{\rm s}-2)})^2 
\right]^{\frac{1}{2}}/E^{(N_{\rm s}-2)}, \nonumber
\end{eqnarray}
where $N_{\rm s}$ indicates the beginning of the saturation region. In our 
special example, $N_{\rm s}=7$ (see Figure \ref{conv}.$a$).
In Eq. (\ref{Epms}) the value of the coupling constant $g$ is chosen such
that the error of the resummation becomes small and the influence
of $\alpha$ is isolated. Since the rate of convergence of the resummation
decreases for increasing $g$, we determine the 
optimal $\alpha$ at a small coupling constant $g/4=0.1$.
In Figure \ref{conv}.$b$, we have plotted the $\alpha$-dependence of
$\Delta^{N_{\rm s}}(g,\alpha)$ for $N_{\rm s}=7$ and $g/4=0.1$. The optimal
$\alpha$-value is found to be ${\alpha}^{\Delta}_{\rm PMS}=0.341$,
which is very close to the exactly known 
strong-coupling parameter $\alpha=1/3$.

\begin{figure}[p]
\unitlength1cm
\vspace{-8ex}
\begin{center}
\begin{minipage}[t]{10.2cm}
\hfill
\epsfxsize=10cm
\epsfbox{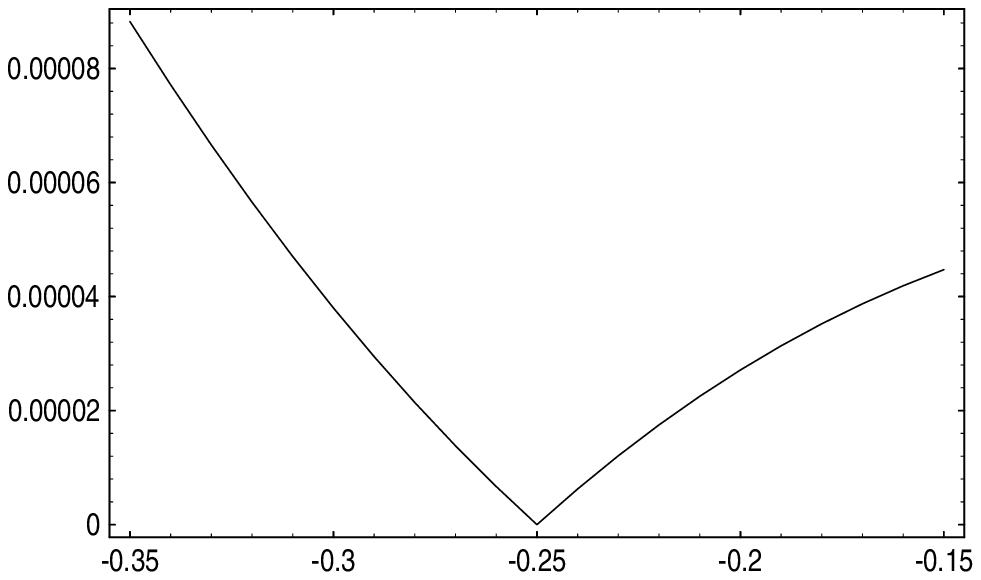} \par 
\begin{picture}(10,2)
\put(0.1,7.5){$\Delta^{(7)}$}
\put(2.1,5.5){$g/4$ = $0.1$}
\put(4.2,8.0){${\alpha}^{\Delta}_{\rm min}$ = ${\alpha}_{\rm exact}$ = $-1/4$}
\put(5.7,3.7){$\alpha$}
\end{picture}
\end{minipage}
\end{center}
\vspace{-20ex}  
\caption[]{\label{Iconv} 
Judicial function $\Delta^{(7)}(g,\alpha)$ 
of Eq. (\ref{Epms}) to determine the strong-coupling power $\alpha$
for the simple integral $Z(g)=\int dx \exp(-x^2/2-gx^4/4)$, plotted as a
function of $\alpha$ for $g/4=0.1$. The optimal  
parameter $\alpha^{\Delta}_{\rm min}$ lies at the lowest value of
$\Delta^{(7)}(g,\alpha)$, and is found to be equal to the exact one. 
It can be shown that for $\alpha=\alpha_{\rm exact}=-1/4$, already 
the zeroth order of the resummation algorithm used in this paper 
reproduces the exact value of $Z(g)$ \cite{KleJan}. 
This is the origin of the cusp and $\Delta^{(N_{\rm s})}(g,\alpha)=0$ 
at $\alpha=-1/4$.}
\end{figure}

We mention that for the zero-dimensional case which corresponds to a
simple integral with $g x^4/4$-interaction, the above criterion 
yields the exact strong-coupling parameter $\alpha=-1/4$
(see Figure \ref{Iconv}).
  
In most applications of quantum field theory, perturbation series
are too short to detect the formation of the saturation plateau with
sufficient accuracy. We shall see in the following that for shorter
series the criterion given above can be simplified without a significant 
loss of accuracy by neglecting the curvature term in Eq. (\ref{Epms}). Then
the slope term is deduced from the two highest known partial sums
$E^{(N_{\rm max}-1)}$ and $E^{(N_{\rm max})}$ by forming:
\begin{equation}
\label{Deltappr}
\Delta^{(N_{\rm max}+1)}(g,\alpha) \approx
\frac{E^{(N_{\rm max})}(g,\alpha)}{E^{(N_{\rm max}-1)}(g,\alpha)}-1
\end{equation} 
where it is important that $E^{(N_{\rm max}-1)}$ and $E^{(N_{\rm max})}$ 
have gotten over the large-fluctuating initial region. Thus the optimal 
$\alpha$-value, denoted now by ${\alpha}_{\rm PMS}$, will be determined
from calculating the extrema or turning points of 
$E^{(N_{\rm max})}(g,\alpha)/E^{(N_{\rm max}-1)}(g,\alpha)$. 
As before, the coupling constant is chosen to be small $g/4=0.1$ to
ensure a sensitive determination of the optimal $\alpha$.
  
\begin{figure}[p]
\unitlength1cm
\vspace{-8ex}
\begin{minipage}[t]{13.2cm}
\epsfxsize=13cm
\hfill
\epsfbox{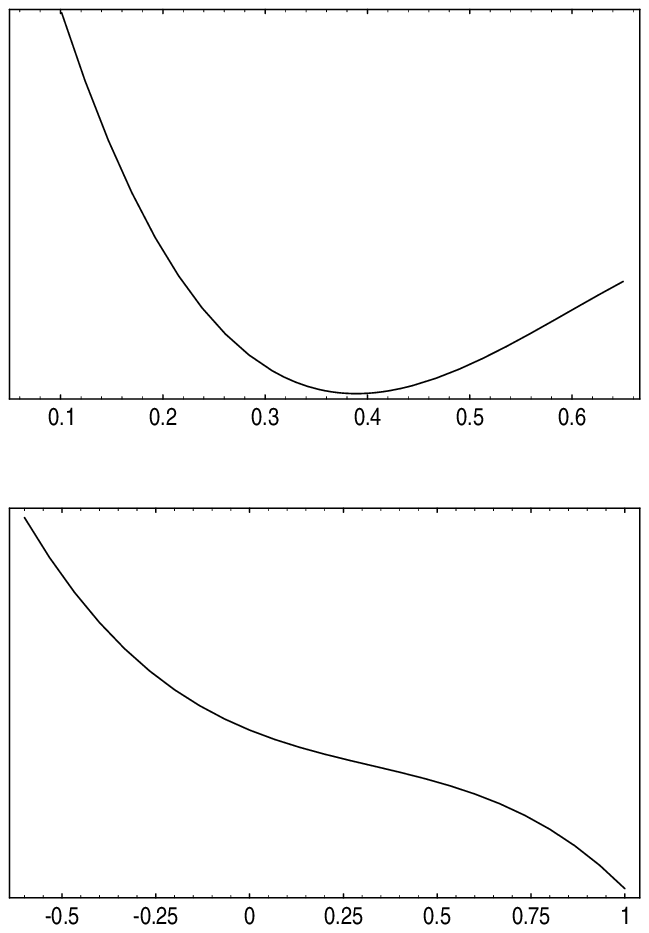} \par
\begin{picture}(13,2)
\put(0.65,14.4){{\small 1.00001075}}
\put(0.5,12.5){$E^{(6)}/E^{(5)}$}
\put(3.0,10.3){$g/4$ = $0.1$}
\put(1.0,9.7){{\small 1.000007}}
\put(1.2,7.9){{\small 1.00015}}
\put(0.5,6.0){$E^{(6)}/E^{(5)}$}
\put(1.2,3.15){{\small 0.99995}}
\put(6,14.0){$a)$\ $M$ = $1$ }
\put(6,12.4){${\alpha}_{\rm PMS}$ = $0.389$ }
\put(6,13.0){${\alpha}_{\rm exact}$ = $1/3$}
\put(6,7.5){$b)$\ $M$ = $2$ }
\put(6,6.5){${\alpha}_{\rm exact}$ = $1/3$}
\put(6,5.9){${\alpha}_{\rm PMS}$ = $0.323$ }
\put(6.7,8.85){$\alpha$}
\put(6.7,2.25){$\alpha$}
\end{picture}
\end{minipage}
\vspace{-10ex}
\caption[]{\label{e0M12} 
The $\alpha$-dependence of the ratio $E^{(6)}/E^{(5)}$
in Eq. (\ref{Deltappr}) for the
O($M$)-symmetric anharmonic oscillator at constant coupling strength
$g/4=0.1$ and various numbers of field components $M$.
$a)$ For the simple anharmonic oscillator $(M=1)$
the optimal $\alpha$-value ${\alpha}_{\rm PMS} = 0.389$ lies at the minimum.
$b)$ For the two-component oscillator ($M=2$) the optimal value 
${\alpha}_{\rm PMS} = 0.323$ lies at the turning point.}
\end{figure}

As the largest available order of approximation for the RG-functions in
QFT is $N_{\rm max}=6$, we have plotted in
Figure \ref{e0M12} the $\alpha$-dependence
of the ratio $E^{(6)}/E^{(5)}$ for the numbers of vector components $M=1$ and
$M=2$ at a coupling constant $g/4=0.1$. In the case of $M=1$, 
a minimum exists at ${\alpha}_{\rm PMS}=0.389$. 
For $M=2$, there is no extremum, and the
optimal $\alpha$-value lies at the turning point ${\alpha}_{\rm PMS}=0.323$.
In both cases the simplified criterion yields again results for
${\alpha}_{\rm PMS}$ which are close to the exactly known 
strong-coupling parameter $\alpha=1/3$.

It is unnecessary to know $\alpha$ to a higher accuracy than that.
In order to show this, we have compared the resummed
ground-state energies $E^{(N)}$ of the anharmonic oscillator
using various values of $\alpha$ (see Table \ref{Te0alph}).
For the coupling constants $g/4=0.1$ and $g/4=1.0$, 
the rate of convergence to the exact energies depends only very
little on $\alpha$.

\begin{figure}[p]
\unitlength1cm
\vspace{-8ex}
\begin{minipage}[t]{13.2cm}
\epsfxsize=13cm
\hfill
\epsfbox{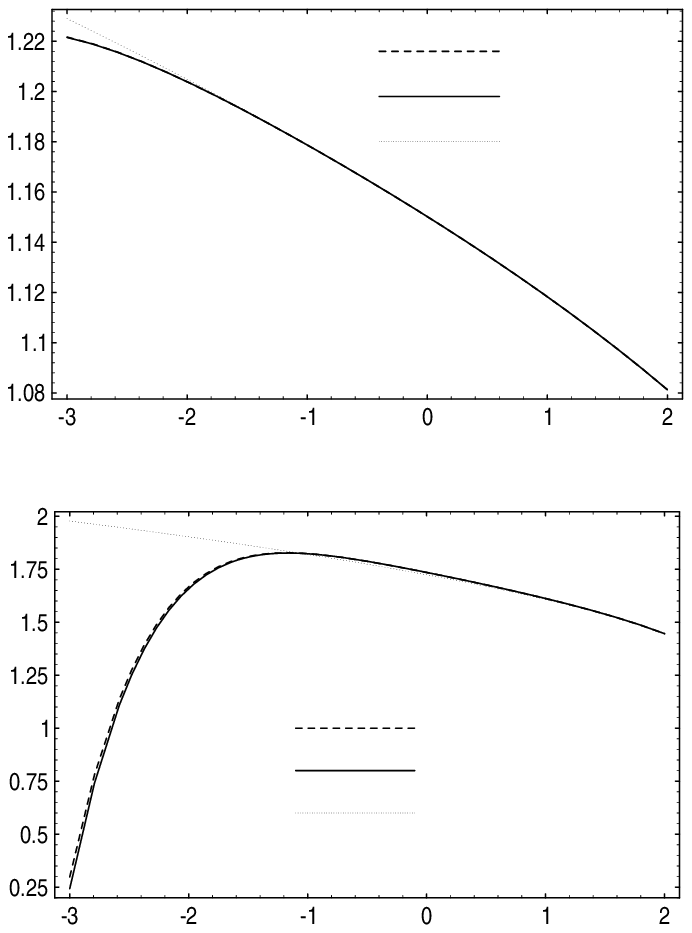} \par
\begin{picture}(13,2)
\put(1.5,12.5){$E^{(6)}$}
\put(1.5,6.0){$E^{(6)}$}
\put(4.0,11.0){$g/4$ = $0.1$ }
\put(9.4,14.2){${\alpha}_{\rm PMS}$ }
\put(9.4,13.6){${\alpha}_{\rm exact}$ }
\put(9.4,13.0){VPT}
\put(6.3,6.5){$g/4$ = $1.0$}
\put(8.3,5.4){${\alpha}_{\rm PMS}$ }
\put(8.3,4.85){${\alpha}_{\rm exact}$ }
\put(8.3,4.30){VPT}
\put(6.7,8.85){$\delta$}
\put(6.7,2.25){$\delta$}
\end{picture}
\end{minipage}
\vspace{-10ex}
\caption[]{\label{PMS6} 
Ground state energy $E$ of
the anisotropic oscillator with the interacting
potential (\ref{QMkub}), as a function of the anisotropy parameter $\delta$
for two coupling constants $g/4=0.1$ and $g/4=1.0$.
Comparison is made between the approximation $E^{(6)}$ obtained once for the
exact strong-coupling parameter ${\alpha}_{\rm exact}=1/3$ and once for
${\alpha}_{\rm PMS} = 0.323$. In the case $g/4=0.1$, differences are 
invisible at this graphical resolution. 
For comparison, results from another resummation scheme,
variational perturbation theory (VPT)~\cite{KleiPI}, obtained 
in Ref.~\cite{KThoJa} are shown as well.}
\end{figure}

Consider now the case that $Z(g,\delta)$ in Eq. (\ref{Zseries}) 
represents the ground-state energy
$E(g,\delta)$ of the anisotropic quartic oscillator with an interaction
\begin{equation}
\label{QMkub}
V_{int}=\frac{g}{4}\left[x^4+2(1-\delta)x^2 y^2+y^4\right]
\end{equation}
which we have studied in detail in \cite{KThoJa} using the Borel-type
resummation algorithm of Refs. \cite{ResPro,KleJan}. 
When expressed in terms of the
old coupling constants $u$, $v$ via $g=u+v$ and $\delta=v/(u+v)$, the
expression (\ref{QMkub}) corresponds to the interaction term in the
euclidean action (\ref{Hamilt}) for the number of field components $M=2$.
It has been found in \cite{KThoJa} that the parameter $\alpha$ is the same for
each coefficient $E_n(g)$ in the $\delta$-expansion of the ground-state energy
$E(g,\delta)=\sum_{n=0}^{\infty} E_n(g) {\delta}^n$,
and having at each $n$ the same value $\alpha=1/3$ as in the isotropic
case. The approximation $E^{(N)}_n(g)$ which
follows from a resummation of the corresponding perturbation series in $g$
up to the order $N$
is becoming less accurate for increasing $n$ since the $g$-expansions
have fewer and fewer terms. However, taking into account the smallness of the
anisotropy $\delta$, the $\delta$-expansion may be truncated at a finite order
$N$. This yields the approximation
\begin{displaymath}
E^{(N)}(g,\delta)= \sum_{n=0}^N E^{(N)}_n(g) {\delta}^n,
\end{displaymath}
which was found to be very accurate in a wide region of $\delta$
around $\delta=0$.
We have compared the result for the ground-state energy resummed at
$\alpha=1/3$, which is known from Ref. \cite{KThoJa},
with the result obtained for ${\alpha}_{\rm PMS}=0.323$.
Figure \ref{PMS6} shows the
$\delta$-dependence of the approximated ground-state energy
$E^{(6)}(g,\delta)$ for the two different values of the parameter $\alpha$
and various coupling constants $g/4$. For $g/4=0.1$, the two curves
for $\alpha$ and ${\alpha}_{\rm PMS}$ coincide. From Figure \ref{PMS6}
and Table \ref{Te0alph} we can thus conclude that the error which 
is caused by an inaccurate determination of $\alpha$ is negligible.
\subsection{Application to Quantum Field Theory}
The above analysis of the anisotropic oscillator can now be applied
to the corresponding model in quantum field theory.
The resummation of the $\beta$-functions (\ref{zeros}) is carried out
by combining the formulas (\ref{Zapprox}), (\ref{ZIpn}), and (\ref{apn}), where
the function $Z$ stands now for ${\beta}^u$ and ${\beta}^v$. The parameters
$b_0(n)$ and $\sigma$ follow from the large-order behavior (\ref{betaLO}), and
are the same for both $\beta$-functions:
\begin{eqnarray}
b_0(n)\!\!\!&=&\!\!\!{\beta}_n+\frac{3}{2}=6+n \, , \nonumber \\
\sigma \!\!\!&=&\!\!\! 1 \, .
\end{eqnarray}
The optimal value of the parameter $\alpha$ is chosen to cause minimal
sensitivity of the ratios $B^{u(6)}_{0}/B^{u(5)}_{0}$ and 
$B^{v(6)}_{1}/B^{v(5)}_{1}$ on $\alpha$ at the small value of coupling
constant $g/4=0.1$ [recall Eq. (\ref{zeros})] in accordance with the above
observations for the O($M$)-symmetric anharmonic oscillator.

The optimal values ${\alpha}_{\rm PMS}$ are found to be:
\begin{equation}
{\beta}^u :\quad {\alpha}_{\rm PMS}=1.348 \quad , \quad \quad
{\beta}^v :\quad {\alpha}_{\rm PMS}=1.225 \, .
\end{equation}
For the simultaneous solution of Eqs. (\ref{zeros}), we have first
determined for each $\beta$-function all zero-point
functions ${\delta}^{(i)}(g)$ which are implicitly defined by
\begin{equation}
\label{deltIMPL}
{\beta}^{u/v}[g,{\delta}^{(i)}_{u/v}(g)]=0 \, .
\end{equation}
From the second equation in (\ref{zeros}), we have then read off a
trivial solution ${\delta}^{(1)}_v(g) \equiv 0$. 
Restricting attention to the region around the isotropic limit $\delta=0$, 
we have found numerically a second nontrivial solution ${\delta}^{(2)}_v(g)$. 
For ${\beta}^u$ only one solution ${\delta}_u(g)$ was found. 
Having obtained ${\delta}^{(1)}_v$ and ${\delta}_u$ the isotropic
fixed point follows from the condition
\begin{equation}
{\delta}^{\ast}_{u \, {\rm iso}}(g^{\ast}_{\rm iso})=
{\delta}^{\ast (1)}_{v \, {\rm iso}}(g^{\ast}_{\rm iso}) \equiv 0 \, .
\end{equation}
The cubic fixed point is similarly obtained by calculating the solution of
the equation:
\begin{equation}
{\delta}^{\ast}_{u \, {\rm cub}}(g^{\ast}_{\rm cub})=
{\delta}^{\ast (2)}_{v \, {\rm cub}}(g^{\ast}_{\rm cub}) \, .
\end{equation}
\section{Results}
\begin{figure}[p]
\unitlength1cm
\vspace{-8ex}
\begin{minipage}[t]{13.2cm}
\epsfxsize=13cm
\hfill
\epsfbox{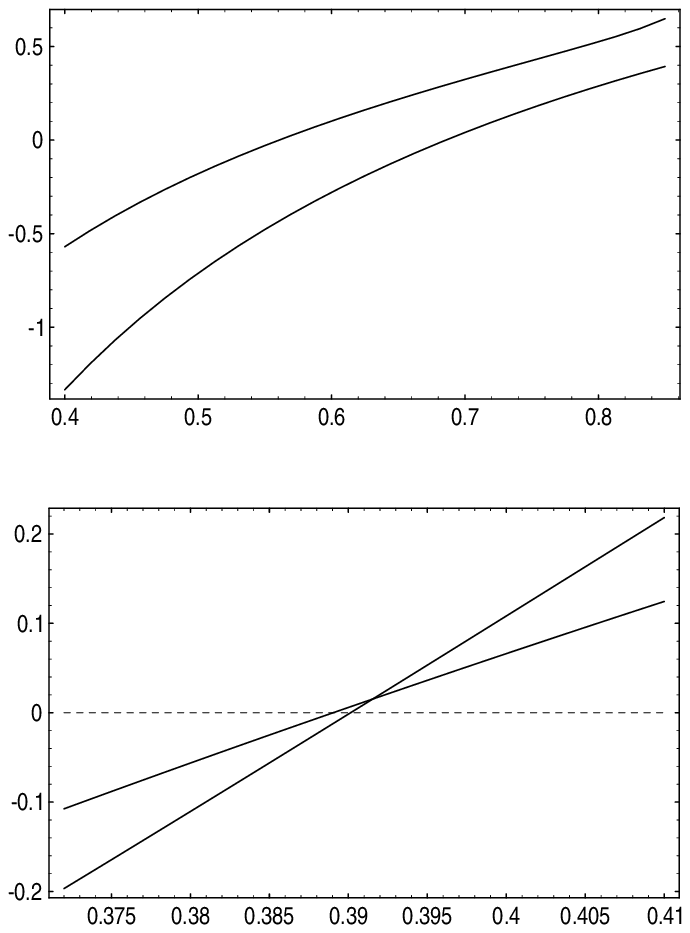} \par
\begin{picture}(13,2)
\put(1.8,13.0){${\delta}_u$}
\put(1.8,12.0){${\delta}_v^{(2)}$}
\put(1.8,6.5){${\delta}_u$}
\put(1.8,5.5){${\delta}_v^{(2)}$}
\put(4.0,14.0){$a)$ }
\put(8.3,10.5){$N$ = $2$}
\put(4.0,7.5){$b)$}
\put(8.3,4.30){$N$ = $6$}
\put(6.7,8.85){$g$}
\put(6.7,2.25){$g$}
\end{picture}
\end{minipage}
\vspace{-10ex}
\caption[]{\label{plot26} 
Determination of the cubic fixed point
$g^{\ast}$, ${\delta}^{\ast}$. The nontrivial
zeros ${\delta}_u$ and ${\delta}_v^{(2)}$ of the $\beta$-functions
${\beta}^u$ and ${\beta}^v$ are plotted
against $g$ [see Eqs. (\ref{zeros}) and (\ref{deltIMPL})].
The the values of $g^{\ast}$ and ${\delta}^{\ast}$ are found from the 
intersection point.
$a)$ Up to the order of approximation $N=2$, no cubic fixed point exist.
$b)$ For the order of approximation $N=6$, the cubic fixed point is
given by $g^{\ast}=0.39154$, ${\delta}^{\ast}=0.015309$. }
\end{figure}

Table \ref{TabCFP} contains the numerical values of the isotropic and the cubic
fixed points for increasing order of approximation $N$. Starting from the
order $N=3$ a cubic fixed point is obtained which lies in the upper half 
of the plane of coupling constants $u$ and $v$ $(\delta > 0)$.
Plotting the functions ${\delta}_u(g)$ and ${\delta}^{(2)}_v(g)$, 
yields the cubic fixed point $(g^{\ast}_{\rm cub},{\delta}^{\ast}_{\rm cub})$
via the crossing point. This is shown in Figure \ref{plot26} for the orders
of approximation $N=2$ and $N=6$.
The cubic fixed point is found to lie very close to the isotropic one.

In order to convince ourselves of the stability of the cubic fixed point 
at the number of field components $M=3$,
we calculate the eigenvalues $b_1$ and $b_2$ of the matrix
\begin{equation}
\label{Eigw}
B=\left. \left(
\begin{array}{cc}
 {\partial}_g {\beta}^g & {\partial}_{\delta} {\beta}^g \\
 {\partial}_g {\beta}^{\delta} & {\partial}_{\delta} {\beta}^{\delta}\\
\end{array} \right) \right|_{g^{\ast},{\delta}^{\ast}}
\end{equation}
using the resummed $\beta$-functions (\ref{Beta}).
The result is contained in Table \ref{Eigwert}.
If both eigenvalues are positive, the corresponding fixed point is infrared
stable. For $M=3$, this is definitely the case for the cubic fixed point.
At the isotropic fixed point, one the other hand,
one eigenvalue $b_2$ is negative.
As the isotropic and the cubic fixed point interchange their stability
at $M=M_c$, the result corroborate the suggestion
in \cite{MaSo} and \cite{KleSchu2} that the
critical value $M_c$ lies below $M=3$. 

Thus we conclude that
the critical behavior of magnetic phase transitions in
anisotropic crystals with cubic symmetry is governed by the cubic,
not by the isotropic Heisenberg fixed point.
The corresponding critical exponents $\eta$ and $\nu$ follow from
the resummed RG-functions ${\gamma}_{\phi}$ and ${\gamma}_{{\phi}^2}$ via the
defining relations
\begin{equation}
\eta={\gamma}_{\phi}(g^{\ast},{\delta}^{\ast}) \, , \quad
{\nu}^{-1}-2=-{\gamma}_{{\phi}^2}(g^{\ast},{\delta}^{\ast}).
\end{equation}
Unfortunately, our result is only of fundamental interest
and has no easily measurable experimental consequences.
Due to the vicinity of the isotropic fixed point, the difference in the
critical exponents is smaller than one percent, so that the
new cubic universality class is practically indistinguishable from the 
isotropic class.

\newpage
\begin{center}\bf F\ i\  g\ u\ r\ e\ \ \ C\ a\ p\ t\ i\ o\ n\ s \end{center}
\vspace*{1.5cm}
{\bf Figure 1:} Convergence of the ground-state energy $E$ of the anharmonic
oscillator with $g/4=0.1$. $a)$ Resummed ground-state energies $E^{(N)}$ are
plotted versus the order of approximation {\it N} for various
strong-coupling parameters $\alpha$. $b)$ The $\alpha$-dependence of the 
function $\Delta^{(7)}$ of Eq. (\ref{Epms}).
The optimal $\alpha$-value is given by the 
minimum ${\alpha}^{\Delta}_{\rm PMS}=0.3408$, very close to the exactly
known value $1/3$.

\vspace*{1cm}
\noindent
{\bf Figure 2:} Judicial function $\Delta^{(7)}(g,\alpha)$ 
of Eq. (\ref{Epms}) to determine the strong-coupling power $\alpha$
for the simple integral $Z(g)=\int dx \exp(-x^2/2-gx^4/4)$, plotted as a
function of $\alpha$ for $g/4=0.1$. The optimal  
parameter $\alpha^{\Delta}_{\rm min}$ lies at the lowest value of
$\Delta^{(7)}(g,\alpha)$, and is found to be equal to the exact one. 
It can be shown that for $\alpha=\alpha_{\rm exact}=-1/4$, already 
the zeroth order of the resummation algorithm used in this paper 
reproduces the exact value of $Z(g)$ \cite{KleJan}. 
This is the origin of the cusp and $\Delta^{(N_{\rm s})}(g,\alpha)=0$ 
at $\alpha=-1/4$.   

\vspace*{1cm}
\noindent
{\bf Figure 3:} The $\alpha$-dependence of the ratio $E^{(6)}/E^{(5)}$
in Eq. (\ref{Deltappr}) for the
O($M$)-symmetric anharmonic oscillator at constant coupling strength
$g/4=0.1$ and various numbers of field components $M$.
$a)$ For the simple anharmonic oscillator $(M=1)$
the optimal $\alpha$-value ${\alpha}_{\rm PMS} = 0.389$ lies at the minimum.
$b)$ For the two-component oscillator ($M=2$) the optimal value 
${\alpha}_{\rm PMS} = 0.323$ lies at the turning point.

\vspace*{1cm}
\noindent
{\bf Figure 4:} Ground state energy $E$ of
the anisotropic oscillator with the interacting
potential (\ref{QMkub}), as a function of the anisotropy parameter $\delta$
for two coupling constants $g/4=0.1$ and $g/4=1.0$.
Comparison is made between the approximation $E^{(6)}$ obtained once for the
exact strong-coupling parameter ${\alpha}_{\rm exact}=1/3$ and once for
${\alpha}_{\rm PMS} = 0.323$. In the case $g/4=0.1$, differences are 
invisible at this graphical resolution. 
For comparison, results from another resummation scheme,
variational perturbation theory (VPT)~\cite{KleiPI}, obtained 
in Ref.~\cite{KThoJa} are shown as well.

\vspace*{1cm}
\noindent
{\bf Figure 5:} Determination of the cubic fixed point
$g^{\ast}$, ${\delta}^{\ast}$. The nontrivial
zeros ${\delta}_u$ and ${\delta}_v^{(2)}$ of the $\beta$-functions
${\beta}^u$ and ${\beta}^v$ are plotted
against $g$ [see Eqs. (\ref{zeros}) and (\ref{deltIMPL})].
The the values of $g^{\ast}$ and ${\delta}^{\ast}$ are found from the 
intersection point.
$a)$ Up to the order of approximation $N=2$, no cubic fixed point exist.
$b)$ For the order of approximation $N=6$, the cubic fixed point is
given by $g^{\ast}=0.39154$, ${\delta}^{\ast}=0.015309$.
\newpage
%
%
\newpage
%
%
\newpage
%
%
\newpage
%
%
\newpage
%
%
\newpage
%
%
\begin{table}[p]
\begin{center}
\begin{tabular}{|c|ll|ll|} \hline \hline
 $E_0$ &\multicolumn{2}{c|}{$g/4=0.1$}&
        \multicolumn{2}{c|}{$g/4=1.0$} \\ \hline
     $N$ & \hspace{0.5em} ${\alpha}_{\rm exact}$ &
       \hspace{0.5em} ${\alpha}_{\rm PMS}$ &
       \hspace{0.5em} ${\alpha}_{\rm exact}$ &
       \hspace{0.5em} ${\alpha}_{\rm PMS}$ \\ \hline

     1 & 0.561496 & 0.56235 & 0.83055 & 0.849631 \\

     2 & 0.558592 & 0.558614 & 0.78297 & 0.784942 \\

     3 & 0.559232 & 0.559254 & 0.812948 & 0.816638 \\

     4 & 0.559142 & 0.559142 & 0.801761 & 0.802012 \\

     5 & 0.559143 & 0.559143 & 0.802206 & 0.802487 \\

     6 & 0.559147 & 0.559147 & 0.805103 & 0.805518 \\

     7 & 0.559146 & 0.559146 & 0.803901 & 0.803937 \\

     8 & 0.559146 & 0.559146 & 0.803115 & 0.803072 \\

     9 & 0.559146 & 0.559146 & 0.803852 & 0.803924 \\ \hline
 exact & \multicolumn{2}{c|}{0.559146}& \multicolumn{2}{c|}{0.803770}
\\ \hline \hline
\end{tabular}
\caption{\label{Te0alph} Convergence of the ground-state energy $E_0$ of
the anharmonic oscillator with $M=1$ for the strong-coupling parameters
${\alpha}_{\rm exact}=1/3$ and ${\alpha}_{\rm PMS}=0.389$.}
\end{center}
\end{table}
%
%
\newpage
\begin{table}[p]
\begin{center}
\begin{tabular}{|c|lc|ll|} \hline \hline
   $N$ & \hspace{1em} $g^{\ast}_{\rm iso}$ &
        ${\delta}^{\ast}_{\rm iso}$ &
       \hspace{1em} $g^{\ast}_{\rm cub}$ &
       \hspace{1em} ${\delta}^{\ast}_{\rm cub}$ \\ \hline

   2 & 0.560616 & 0 & \multicolumn{2}{c|}{don't exist}  \\

   3 & 0.440796 & 0 & 0.50208  & 0.291074 \\

   4 & 0.393506 & 0 & 0.400199 & 0.037862 \\

   5 & 0.4012 & 0 & 0.411057 & 0.063068 \\

   6 & 0.389037 & 0 & 0.39154  & 0.015309 \\ \hline \hline
\end{tabular}
\caption{\label{TabCFP} Numerical result for the isotropic and the cubic
fixed point for increasing order of approximation $N$. For $N \geq 3$,
a cubic fixed point is found in the upper half of the
coupling constant plane $(u,v)$,
i.\ e.\ ${\delta}^{\ast}_{\rm cub} > 0$.}
\end{center}
\end{table}
\newpage
%
%
\begin{table}[p]
\begin{center}
\begin{tabular}{|c|ll|ll|} \hline \hline
   $N$ & \hspace{1em} $b_1^{\rm cub}$ &
         \hspace{1em} $b_2^{\rm cub}$ &
         \hspace{1em} $b_1^{\rm iso}$ &
         \hspace{1em} $b_2^{\rm iso}$ \\ \hline

   4 & 0.782796 & 0.0048920  & 0.784532 & -0.00502046  \\

   5 & 0.764835 & 0.00851725 & 0.763966 & -0.00886277  \\

   6 & 0.80609  & 0.00212717 & 0.80658  & -0.00214788  \\ \hline \hline
\end{tabular}
\caption{\label{Eigwert} Stability of cubic fixed point, as demonstrated
by the eigenvalues of the stability matrix $B$
in Eq. (\ref{Eigw}), calculated from the resummed $\beta$-functions
${\beta}_g$ and ${\beta}_{\delta}$ in Eq. (\ref{Beta}).
The isotropic fixed point is unstable.}
\end{center}
\end{table}
\end{document}